# UNA CONTROVERSIA HISTÓRICA AL SERVICIO DE UNA SITUACIÓN DE APRENDIZAJE: UNA RECONSTRUCCIÓN DIDÁCTICA BASADA EN *DIÁLOGO SOBRE LOS DOS MÁXIMOS SISTEMAS DEL MUNDO* DE GALILEO[1]

de Hosson, Cécile
Laboratoire de Didactique André Revuz, Université Paris Diderot
cecile.dehosson@univ-paris-diderot.fr

**Resumen.** Para el sentido común resulta difícil concebir que, al soltar un objeto desde la cima del mástil de una nave que avanza a velocidad constante, éste cae al pie del mástil porque conserva su movimiento horizontal. Esta dificultad nos evoca aquella que acechó a los sabios de la ciencia preclásica, y que Galileo puso en escena en su *Diálogo sobre los dos máximos sistemas del mundo*. De allí nace una trayectoria de aprendizaje en la cual se seleccionan y se reorganizan los elementos del *Diálogo* según su pertinencia didáctica. La eficacia de la trayectoria, evaluada en el marco de la ingeniería didáctica, se debe en parte a la identificación de los alumnos con los personajes que Galileo pone en escena.

**Palabras clave.** Controversia, historia de las ciencias, Galileo, conservación, ingeniería didáctica.

**Using a historical controversy in a learning context: the case of a «didactic engineering» elaborated from Galileo's *Dialogue Concerning The Two Chief World Systems***

**Summary.** It is difficult for common sense to admit that an object dropped from the top of the mast of a ship moving at a constant velocity falls down at the bottom of the mast because it keeps within the horizontal movement of the ship. This difficulty is similar to the one faced by early scientists from the pre-classical science and staged by Galileo in his *Dialogue Concerning The Two Chief World Systems*. This proximity leads us to elaborate a learning pathway in which some elements of Galileo's dialogue are selected and reorganized according to specific educational constraints. The relevance of the teaching-learning sequence is asserted by the 'didactic engineering' framework and leans on the identification of students with the characters staged by Galileo.

**Keywords.** Controversy, History of Science, conservation, Galileo, Didactic Engineering.

## INTRODUCCIÓN

Introducir elementos de historia de las ciencias en la enseñanza de las ciencias es un requisito presente en varios currículos científicos. En Francia, el objetivo principal de la enseñanza de las ciencias en el último ciclo del bachillerato es «despertar en los alumnos el amor por la ciencia». Para lograrlo, se pretende que los alumnos comprendan «el proceso intelectual, la evolución de las ideas, la construcción progresiva del corpus de conocimiento científico» (Anexo del 2.º número especial del BOEN –*Bulletin Officiel de l'Éducation Nationale*–, 31 de agosto de 2001). Como testigo de las preguntas, las controversias, las contingencias que jalonan la génesis de un descubrimiento, la historia de las ciencias se presenta con frecuencia como un antídoto contra el dogmatismo de la clase de ciencias tradicional.

No es fácil involucrar a los alumnos en la dinámica compleja de un descubrimiento científico. Ello requiere que los soportes potenciales (texto original, documento iconográfico, relato histórico, réplica de experiencias…) estén adaptados a los objetivos pedagógicos y didácticos asignados. El reto consiste en evitar, en la medida de lo posible, el estilo reductor y anecdótico utilizado, por ejemplo, en muchos manuales escolares, cuya eficacia resulta discutible en términos de aprendizaje (Mathy, 1997; Slisko, 2008). Por otro lado, independientemente del soporte que se elija, es importante que éste se inscriba en una secuencia que favorezca la actividad de investigación según un acercamiento constructivista. En esta perspectiva, la estrategia que consiste en transponer un problema histórico a la clase de ciencias parece particularmente productiva (Monk y Osborne, 1997; Matthews, 2000; Galili y Hazan, 2001; de Hosson y Kaminski, 2007). El procedimiento consiste en elegir una pregunta que suponemos pueda propiciar un debate dentro de la clase, y concebir una trayectoria cognitiva que se apoye en dicho debate y que conduzca a los alumnos a la construcción de un conocimiento científico determinado. Utilizar así la historia de las ciencias requiere un conocimiento previo y creciente del estado conceptual de los alumnos, en particular de las ideas y de los razonamientos que pueden obstaculizar el aprendizaje previsto. También es necesario que la pregunta histórica tenga sentido en la clase para que los alumnos se la puedan apropiar más fácilmente y para suscitar la emergencia de propuestas de solución. La elección del problema, que se tomará de la ecología histórica, está por tanto sujeta a las coacciones que fundamentan la ecología didáctica (dificultades relacionadas con el conocimiento enseñado, ideas previas de los alumnos, tiempo didáctico…).

En este artículo presentaremos una secuencia de enseñanza-aprendizaje de mecánica (noveno grado, alumnos de 15 años) construida a partir del *Diálogo sobre los dos máximos sistemas del mundo*, escrito en 1632 por Galileo. Ésta pretende que los estudiantes aprendan el principio de conservación del movimiento en el contexto de la emergencia del principio de inercia. Intentaremos resaltar algunos rasgos específicos de la elaboración de la secuencia. En particular, insistiremos en la estrategia de selección y organización de los fragmentos del *Diálogo* que serán utilizados. Explicitar estos procedimientos nos conducirá a evaluar el impacto de la secuencia en una situación real de clase.

En primer lugar, evocaremos las dificultades principales asociadas al aprendizaje de la conservación del movimiento (en la física galileo-newtoniana). Este trabajo de exploración didáctica nos permitirá elegir un problema cuya solución en clase requerirá que los alumnos modifiquen su esquema de pensamiento. Con este fin, ellos seguirán las etapas de un proceso cognitivo en el cual se organizarán ciertos elementos históricos (extraídos del *Diálogo sobre los dos máximos sistemas del mundo* de Galileo) para responder a las exigencias del proyecto didáctico que hemos fijado.

## 1. PRESUPUESTOS TEÓRICOS DEL DISEÑO DE LA TRAYECTORIA DE APRENDIZAJE

El diseño, así como el análisis de nuestra secuencia, se inscriben en el marco de la ingeniería didáctica: «El término ingeniería didáctica designa un conjunto de secuencias de clase concebidas, organizadas y articuladas en el tiempo de forma coherente por un profesor-ingeniero para efectuar un proyecto de aprendizaje de un contenido dado para un grupo concreto de alumnos» (Artigue ,1998, 40). De manera cronológica, el proceso experimental de la ingeniería didáctica consta de cuatro fases:

1. Primera fase: Análisis preliminares que incluyen dimensiones epistemológicas (características del saber puesto en funcionamiento), cognitivas (características cognitivas de los alumnos como «conocimientos previos»), didácticas y pedagógicas (características del funcionamiento del sistema de enseñanza).

2. Segunda fase: Concepción y análisis a priori de las situaciones didácticas.

3. Tercera fase: Experimentación.

4. Cuarta fase: Análisis a posteriori y evaluación

Este marco permite medir el impacto de una acción didáctica mediante la confrontación de un análisis a priori –en el cual están implicadas ciertas hipótesis locales subyacentes a las acciones buscadas–, con un análisis a posteriori –basado en los datos que surgen de la realización efectiva de la secuencia. Para llevar a cabo esta investigación hemos asumido tanto el rol de docente como el de investigador.

### 1.1. Dificultades de los alumnos con respecto a los referenciales y a las leyes del movimiento

Aquí se estudian las dimensiones epistemológicas y cognitivas subyacentes al aprendizaje de la conservación del movimiento. Las dificultades de los alumnos en mecánica clásica (dinámica, cinemática) han sido objeto de numerosas exploraciones (Viennot, 1979; Saltiel y Malgrange, 1980; Clement, 1982; Sebastia, 1984; McDermott, 1984; Halloun y Hestenes, 1985) cuyos resultados han sido retomados y sintetizados regularmente en la literatura didáctica (Picquart 2008, Mora & Herrera 2009). Nos interesaremos aquí únicamente en las dificultades de los alumnos relacionadas con la cuestión de la conservación del movimiento. La mayor parte de los alumnos y de los estudiantes piensa que el movimiento de un cuerpo requiere de la acción de un motor. En otras palabras, cuando dos cuerpos en contacto (A y B) están en movimiento, es porque uno de ellos (A) mueve al otro (B). Si se rompe el contacto entre A y B, entonces B pierde instantáneamente la velocidad correspondiente a la acción de A sobre B: «Al comprender el arrastre como una causa del movimiento para el objeto arrastrado, se piensa en consecuencia que la velocidad correspondiente desaparece al mismo tiempo que el vínculo físico» (Viennot, 2002). Este tipo de razonamiento lleva a los alumnos a predecir que, cuando una persona parada sobre un pasillo rodante en movimiento (animado por una velocidad constante) suelta una canica, ésta caerá detrás de la persona (Saltiel y Viennot, 1985). En este caso, la ausencia de vínculo físico con la persona suprime el componente horizontal de la velocidad de la canica; sólo subsiste, en-tonces, el movimiento vertical (Clement, 1982; Halloun y Hestenes, 1985; Harres, 2005). Un razonamiento de estas características no es conforme al principio de conservación: en realidad, la canica conserva el movimiento de la persona incluso cuando ya no está en contacto con ella. Por esta razón (y en ausencia de fricción del aire), la canica cae a sus pies. La concepción de la secuencia que presentaremos a continuación hace referencia a esta dificultad. La secuencia tiene como objetivo ayudar a los alumnos a romper con la idea de que existe una relación causal entre contacto y movimiento, con el fin de acercarse a la idea de conservación.

### 1.2. Características del funcionamiento del sistema didáctico y pedagógico

El marco teórico general de nuestra trayectoria de aprendizaje es constructivista, pues el conocimiento involucrado en la secuencia se considera como una construcción de los alumnos a partir de los esquemas que ya poseen (Ausubel et al., 2000). En esta perspectiva, el aprendizaje supone una construcción que se realiza a través de un proceso mental que conlleva la adquisición de un conocimiento nuevo (aquí: la conservación del movimiento). El marco teórico constructivista nos proporciona una forma de aprender que se contrapone al aprendizaje memorístico y que favorece que el alumno interactúe con el objeto del conocimiento. Esto supone que el material sea potencialmente significativo, es decir, que el nuevo material de aprendizaje pueda relacionarse de manera no arbitraria y sustancial con alguna estructura cognoscitiva específica del alumno (Ausubel et al., 2000, 48). De manera concreta, la conservación del movimiento se transforma en el objeto de un descubrimiento apoyado por una herramienta histórica y surge como resultado de un cambio conceptual (Posner et al., 1982) que desarrollamos a partir de una situación conflictiva: «La idea del cambio conceptual formó parte desde el principio de las aportaciones nucleares del constructivismo. La noción de construcción personal del conocimiento desde las ideas previas de los alumnos supone la necesaria existencia de un cambio conceptual que permita el salto de una concepción a otra. Se ha señalado que en ese cambio conceptual existen varios aspectos clave, entre los que destaca la necesidad de que el que aprende se sienta insatisfecho con sus preconcepciones, de que las nuevas concepciones estén en el ámbito de lo inteligible para él» (Limón, 2004, 93). En esta perspectiva, buscamos un objetivo de aprendizaje que pudiera ser logrado a través de un proceso de cambio conceptual apoyado sobre similitudes entre los obstáculos que encuentran los alumnos y los que debieron ser superados para permitir el surgimiento de las ciencias modernas. Según Gil-Pérez, deben darse rupturas del mismo tipo en las formas de razonamiento de los alumnos para permitirles entrar en una eventual problemática de cambio conceptual (Gil-Pérez et al., 1990).

A esta dimensión constructivista se añade una dimensión social, pues nuestra intención es que la adquisición individual del conocimiento se realice en interacción con los otros, de forma colaborativa. En este marco llamado «constructivismo social» se expone que el ambiente de aprendizaje óptimo es aquel donde existe una interacción dinámica entre los instructores, los alumnos y las actividades que proveen oportunidades para los alumnos de crear su propia verdad, gracias a la interacción con los otros (Doise, 1991). La idea básica del constructivismo social es que las interacciones sociales (intercambios, discusiones entre compañeros) son un motor del progreso cognitivo. En otras palabras, el encuentro entre pares en torno a una tarea puede redundar en avances cognitivos individuales. Así, los intercambios interpersonales pueden llevar a algunos avances cognitivos cuando el alumno se enfrenta a otros puntos de vista distintos al suyo, es decir, cuando el conflicto sociocognitivo se produce durante la interacción. Este conflicto sociocognitivo se define por la heterogeneidad de las opiniones acerca de un problema. La eficacia del conflicto se debe, en primer lugar, a que un enfrentamiento explícito de opiniones diferentes expone en el aula varias respuestas posibles, y en segundo lugar, a que el otro ofrece, además de su respuesta, la información sobre otra línea de razonamiento. Esta confrontación de ideas abre nuevas vías de resolución y proporciona a los alumnos un marco de pensamiento diferente. En este contexto, el objetivo de la herramienta de enseñanza que desarrollamos es un método de enseñanza que Joshua y Dupin llaman «debate científico en la aula» que resulta de un proceso de transposición didáctica (Joshua y Dupin, 2005). Esto implica un cambio en el rol del profesor que se convierte en moderador, coordinador y facilitador del aprendizaje (Tama, 1986). El profesor, en su rol de mediador, debe incorporar objetivos de aprendizaje relativos a las habilidades cognitivas, dentro del currículo escolar y apoyar al alumno para:

• Desarrollar un conjunto de habilidades cognitivas que les permitan optimizar sus procesos de razonamiento.

• Animarle a tomar conciencia de sus propios procesos y estrategias mentales (metacognición) para poder controlarlos y modificarlos (autonomía), mejorando el rendimiento y la eficacia en el aprendizaje.

### 2. CONCEPCIÓN Y ANÁLISIS A PRIORI DE LAS SITUACIONES DIDÁCTICAS

Usualmente, los alumnos de 15 años pueden construir la idea de la conservación del movimiento a partir de la observación de la caída de un objeto soltado por una persona andando con una velocidad constante en bicicleta (o parada sobre un pasillo rodante en movimiento), gracias a una serie de cronofotografías o por medio del *software* que simulan la caída de una piedra. Se ha demostrado que este tipo de actividades permite al alumno alcanzar de manera bastante eficaz el conocimiento requerido, enfrentándose a observaciones inesperadas (Guedj, 2005); sin embargo, la elección que hicimos en esta investigación fue involucrar a los alumnos en una construcción conceptual progresiva sin experimentación, basada en un fragmento del *Diálogo sobre los dos máximos sistemas del mundo* escrito por Galileo Galilei en 1632 donde se exponen los dos experimentos imaginarios: el de la torre y el del barco (Anexo 1). Nuestra intención fue poner a prueba en el aula la estrategia pedagógica utilizada por Galileo en su *Diálogo*, es decir, averiguar cuáles son los efectos de un razonamiento por analogía que reúne dos situaciones de caída libre: la primera se refiere a la caída de una piedra soltada desde la cima de una torre; la segunda se refiere a la caída de una piedra soltada desde la cima del mástil de una nave moviéndose en línea recta a velocidad constante.

En el fragmento presentado en el anexo 1, Galileo (a través del personaje de Salviati) lleva a Simplicio a cambiar su opinión acerca del resultado de las dos caídas libres. El problema planteado por la pregunta del lugar dónde cae una piedra soltada desde la cima (de una torre o del mástil de una nave) es el del movimiento de la Tierra en torno a su eje[1]. Salviati trata de convencer a Simplicio de que «la piedra cae siempre en el mismo lugar de la nave, tanto si está quieta como si se mueve con cualquier velocidad». La hipótesis subyacente a esta afirmación es que el movimiento de la nave se conserva en una piedra soltada desde la cima del mástil de la nave. Esta hipótesis permite a Salviati convencer a Simplicio de que el ejemplo de la caída libre de una piedra soltada de la cima del mástil de una nave no es relevante para demostrar que la Tierra no se mueve, pues el lugar de caída de la piedra es el mismo independientemente de que la nave se mueva o no: «por ser la misma la argumentación referente a la Tierra que a la nave, del hecho de que la piedra caiga siempre perpendicularmente al pie de la torre no se puede inferir nada sobre el movimiento o reposo de la Tierra».

Los razonamientos de los alumnos presentados arriba evocan los que Galileo Galilei pone en escena en el fragmento del anexo 1. Teniendo en cuenta esta proximidad, construimos nuestra secuencia a partir del texto de Galileo. A primera vista, esta elección carece de originalidad: la utilización del *Diálogo* en clase de física se ha convertido en un ritual en la enseñanza de la mecánica, y varias investigaciones se han desarrollado con miras a evaluar su impacto (Kaplan, 1988; Garrison y Lewwill, 1993; Guedj, 2005). La especificidad de nuestro trabajo se relaciona con:

• la explicitación de las elecciones de reconstrucción que presiden la elaboración de la secuencia bajo el marco teórico de la ingeniería didáctica,

• la probación de la capacidad de argumentación de alumnos de 15 años a partir de una situación de conflicto (una controversia),

• la probación de la capacidad de conceptualización de alumnos de 15 años a través de un razonamiento por analogía,

• la medición cualitativa de los efectos del uso de una herramienta histórica sobre la motivación y el desarrollo conceptual de los alumnos.

Nuestra secuencia de enseñanza está construida a partir del fragmento presentado en el anexo 1, y está organizada según se describe en la figura 1. Se apoya en dos hipótesis: 1) que los alumnos se identificarán con el personaje de Simplicio y 2) que reconocerán la equivalencia (para efectos de la caída) del movimiento de la Tierra y de la nave (pasamos por alto los efectos de arrastre y de Coriolis). En efecto, si los alumnos se identifican con las ideas de Simplicio y admiten con él que la piedra soltada desde la cima del mástil de la nave cae detrás del mástil, y si, por otro lado, admiten que lo que ocurre en el barco debería ocurrir también sobre la Tierra (en los dos casos, un objeto es jalado por un motor, y luego desprendido del mismo), entonces constatarán, tal vez, que la transposición de la situación de la nave a la de una piedra soltada desde la cima de una torre conduce a concluir la inmovilidad de la Tierra. Esta situación conflictiva debería conducirlos a construir un razonamiento nuevo que haga compatible la realidad de la experiencia de la nave con la del movimiento de rotación de la Tierra sobre su propio eje. El hecho de haber elegido la rotación de la Tierra en torno a su eje (en lugar de su movimiento de revolución alrededor del Sol) se inspira directamente en la estrategia implementada por Galileo quien **se pone** explícitamente en escena (en la voz de Simplicio) la posición aristotélica acerca del movimiento de la Tierra[2]. Sin embargo, esta elección será discutida en la última parte de este artículo, pues presenta algunos inconvenientes.

La primera etapa de la secuencia consiste en que los estudiantes predigan el lugar de caída de dos piedras: una soltada desde la cima de una torre, y otra soltada desde la cima del mástil de una nave que avanza a una velocidad constante. Esta etapa es la etapa de devolución del problema y debe dar lugar a la aplicación de un debate de aula a través del cual los alumnos argumentan y expresan ideas que pueden ser contradictorias. La segunda etapa busca que los estudiantes admitan la equivalencia de las dos caídas, para poderlos conducir hacia una conclusión incongruente, apoyada en el siguiente razonamiento: si las caídas son equivalentes, entonces la piedra soltada desde la torre debería caer detrás de ella; éste no es el caso, por lo cual la Tierra es inmóvil. Contamos con el efecto producido por esta situación conflictiva (los alumnos de 15 años saben que la Tierra está en movimiento) para que los alumnos contemplen una nueva conclusión respecto al lugar de caída de la piedra soltada desde la cima del mástil (etapa 3) mediante un razonamiento cercano al razonamiento esperado: la piedra conserva la velocidad del barco, incluso cuando se rompe el contacto entre ella y el mástil (etapa 4). El desarrollo de las etapas se justifica por el procedimiento didáctico utilizado por Galileo para enseñar a los «simplicios» de su época el principio de conservación del movimiento (identificación del «hombre de la calle» al personaje de Simplicio, analogía entre dos caídas libres para lograr un conflicto cognitivo, afirmación del hecho de que una piedra soltada del mástil de una nave adelantando a velocidad constante cae al pie del mástil). Sin embargo, en el caso de nuestra secuencia de aprendizaje, el movimiento de la Tierra pasa de ser una incertidumbre a ser una certeza que se considera actualmente como un punto de apoyo para la trayectoria conceptual de los alumnos.

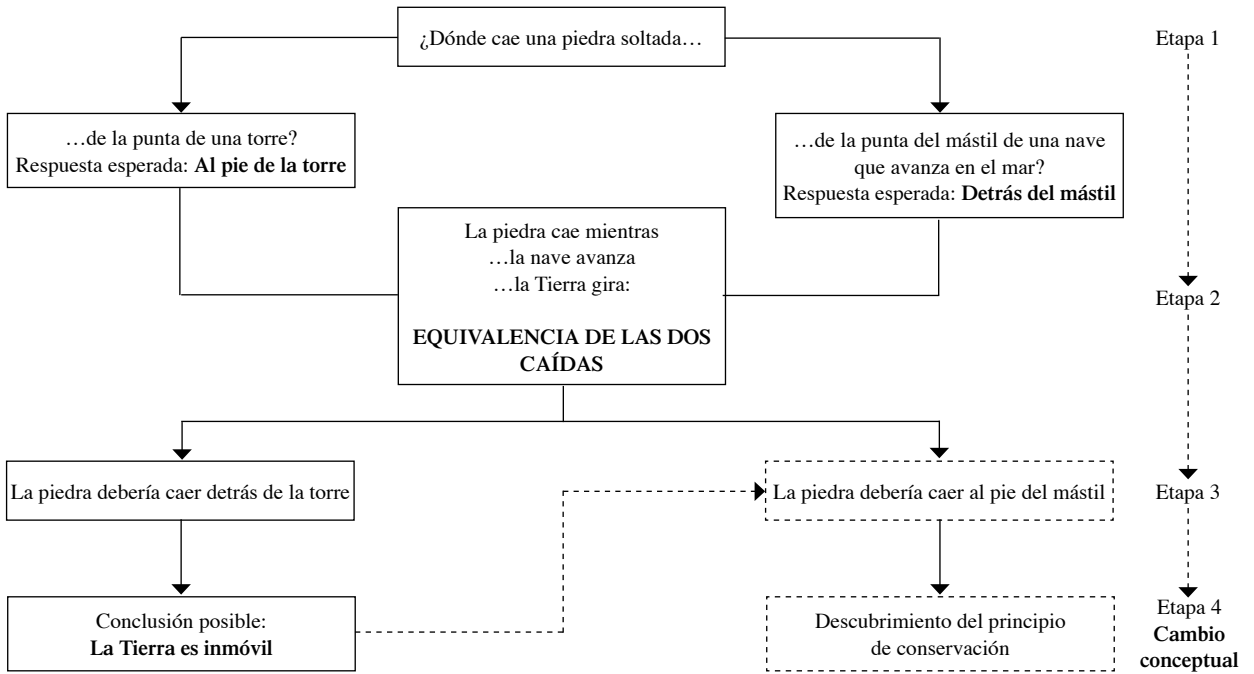

Figura 1
**Organigrama que representa las etapas del recorrido cognitivo propuesto a los alumnos.**

A lo largo de nuestro análisis, buscaremos validar las siguientes hipótesis:

• Identificación de los alumnos con el personaje de Simplicio (H1).

• Efectividad de la asociación de los dos problemas (caída de la piedra en las dos situaciones) (H2).

• Eficacia en términos de cambio conceptual del conflicto cognitivo suscitado por el cuestionamiento del movimiento de la Tierra (H3).

Las hipótesis H1, H2 y H3 se ven justificadas en la tabla 1:

**Tabla 1**

| HIPÓTESIS | JUSTIFICACIONES |
|---|---|
| H1 | Los estudios acerca de los razonamientos de los alumnos de secundaria en mecánica clásica demuestran que la mayor parte de ellos piensa que el movimiento de un cuerpo necesita la acción de un motor. |
| H2 | En ambos casos, la piedra es soltada desde un cuerpo en movimiento |
| H3 | Los alumnos de 15 años saben que la Tierra se mueve en torno a su eje. |

Las reacciones R1, R2 y R3 servirán como indicadores para la validación de las hipótesis H1, H2 y H3 (ver tabla 2). Se trata de una validación cualitativa posterior al análisis de la transcripción de los intercambios entre los alumnos. Los únicos resultados cuantitativos serán los obtenidos gracias a las respuestas escritas al cuestionario presentado en el anexo 2.

Tabla 2
**Correspondencia de las hipótesis y los indicadores que permiten validarlas.**

| HIPÓTESIS | REACCIONES ESPERADAS PARA VALIDAR LA HIPÓTESIS |
|---|---|
| H1 | Los alumnos toman posición dándole de forma explícita la razón a Simplicio, por medio de frases como «estoy de acuerdo con…», «pienso que Simplicio tiene razón»… (R1) |
| H2 | Los alumnos reconocen la equivalencia de las dos caídas y concluyen que la Tierra es inmóvil (R2). |
| H3 | Como consecuencia de R2, los estudiantes cuestionan la respuesta a la pregunta del lugar de caída de la piedra soltada desde la cima del mástil, suponen que ésta caerá al pie del mástil y justifican su respuesta mediante una idea cercana a la de «conservación». (R3) |

## 3. EXPERIMENTACIÓN

Esta secuencia, de una duración total de dos horas, se implementó en tres clases de un liceo de la periferia de París (población socioeconómicamente desfavorecida). 71 alumnos de niveles heterogéneos estuvieron implicados. Se alternaron espacios de intercambio colectivo y tiempos de reflexión individual o en pareja. En una de las clases, la secuencia fue registrada en su totalidad.

Etapa 1: Dónde cae una piedra soltada…

Después de una corta presentación del *Diálogo* y de Galileo, se distribuye a los estudiantes el cuestionario presentado en el anexo 2, y se recoge al cabo de unos 20 minutos. La mayor parte de los alumnos interrogados (96% del total de los alumnos, es decir, 68 sobre 71) prevé que una piedra soltada desde la cima de una torre caerá a sus pies. Sólo una estudiante evoca la idea de que la piedra pueda caer detrás de la torre: «La Tierra gira sobre su propio eje con una velocidad de 260 m/s, entonces deduzco que la piedra caerá detrás de la torre» (Cong). Las respuestas de los estudiantes a la segunda pregunta (el lugar de caída de una piedra soltada desde la cima del mástil de una nave que avanza a una velocidad constante) son conformes a nuestras predicciones. Así, 86% del total de los estudiantes interrogados (61 sobre 71) prevé que la piedra caerá detrás del mástil. La mayoría justifica su respuesta mediante una explicación cercana a la que esperábamos encontrar. Para Franck, «la piedra tocará el suelo detrás, porque cuando el barco avanza, y mientras tanto, la piedra no avanza más, no cambia de trayectoria, cae recto. Yo creo que Simplicio tiene razón». Algunas respuestas están acompañadas por dibujos, como la de Sarah (ver figura 2), quien explica que «en el aire, la piedra cae siempre recto mientras que el barco avanza. Tiene que haber una diferencia de posición de la piedra con respecto al barco. Entonces la piedra cae detrás. Estoy de acuerdo con Simplicio».

Una parte de los alumnos interrogados (11% del total, 8 de 71) parece haber razonado según una intuición galileana, y prevé que la piedra caerá al pie del mástil. Es el caso de Julie, para quien «el barco está en movimiento, entonces la piedra seguirá el movimiento del barco y caerá al pie del mástil, yo creo que es Salviati quien tiene razón», y de Kimnara, que opina que «la piedra caerá al pie del mástil porque cuando se suelta la piedra desde arriba, siempre tiene la velocidad del barco, entonces avanza al mismo tiempo que el mástil. Simplicio se equivoca». Señalemos que 84% de los alumnos interrogados comparan explícitamente su respuesta con la de Simplicio o con la de Salviati de manera coherente. Los demás explicarán luego oralmente que el texto propuesto les pareció muy difícil de entender.

Etapas 2 y 3: Debate y reconocimiento de la equivalencia de las caídas

Al final de la primera etapa recogemos los cuestionarios, y proponemos a los alumnos confrontar sus opiniones mediante un debate abierto iniciado de la siguiente forma:

Prof: Entonces, ¿quién creen ustedes que tiene razón, Simplicio o Salviati? ¿Selim?
Selim: Simplicio, porque la nave sigue avanzando cuando la piedra cae, y la piedra no.
Prof: ¿Por qué la piedra no?
Selim: Porque ya no está agarrada a la nave.
Prof: ¿Todos están de acuerdo con Selim?
Kimnara: No.
Prof: Kimnara, ¿tú qué opinas? ¿Estás de acuerdo con Simplicio?
Kimnara: No, porque la piedra va a la misma velocidad que la nave, incluso cuando ya no está agarrada, como dice Selim, sigue avanzando de todas maneras, al mismo tiempo.

El intercambio dura algunos minutos. Todos los alumnos participan en la discusión, y parecen globalmente muy impacientes por saber quién, si Simplicio o Salviati, tiene razón. Sin embargo, la equivalencia de las dos situaciones no es evocada de forma espontánea por ningún estudiante. Por esta razón, decidimos explicitarla nosotros mismos:

Prof: ¿La Tierra es inmóvil?
La clase: ¡No!
Prof: Pero entonces, si la Tierra gira, es como la nave, ¿no? ¿Por qué la piedra cae detrás del mástil y no detrás de la torre? ¿No es lo mismo?

Figura 2
**Dibujo de Sarah.**

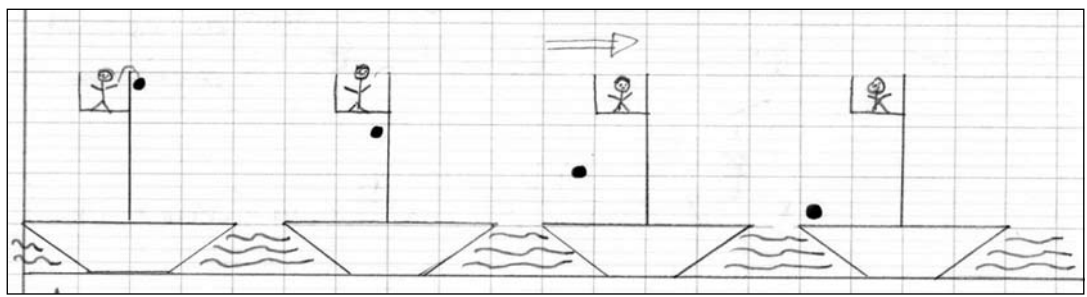

Los alumnos se apropian fácilmente de la pregunta de la equivalencia de las dos situaciones, pero muy rápidamente, ya no comprenden por qué las respuestas que formularon no son idénticas. Para algunos, esto se debe a la diferencia entre la velocidad de desplazamiento de la nave y la de la Tierra:

> Hugo: Es porque la Tierra no gira para nada rápido, de hecho, ni siquiera nos damos cuenta.
> Nordine: Pero claro que sí, ¡gira muy rápido! 260 metros por segundo son casi 1.000 km/h, es mucho más rápido que la nave, entonces la piedra debería aterrizar muy atrás.

Todos los alumnos que respondieron de la misma forma que Simplicio reconocen que su razonamiento los conduce a concluir que la Tierra es inmóvil. Algunos de ellos logran reformular, de forma espontánea, la respuesta que habían propuesto a la pregunta de la nave:

> Nordine: En realidad nos equivocamos sobre la nave, tal vez…
> Prof: ¿Por qué, Nordine?
> Nordine: Porque la Tierra gira y sin embargo la piedra cae al pie de la torre y la nave se mueve, entonces es lo mismo, tiene que haber el mismo razonamiento.
> Elise: Si es lo mismo, tiene que caer al pie del mástil.

La mayoría de los alumnos comprenden que si los dos problemas son equivalentes, entonces la piedra debe caer al pie del mástil. Pero esto les parece tan improbable que se dicen que después de todo, tal vez los problemas no son comparables: «como la Tierra gira en círculos y la nave avanza recto, puede ser que eso explique que la piedra no caiga en el mismo lugar» (Jennifer).

Etapa 4: Construir el principio de conservación

En esta etapa anunciamos a los estudiantes que la predicción correcta es efectivamente la de Salviati Ahora es importante para ellos encontrar una explicación que justifique esta predicción. Les proponemos construir una hipótesis en parejas, diciéndoles que si lo desean pueden hacer dibujos. Algunos retoman los esquemas de las posiciones sucesivas de la nave y ponen la piedra a una distancia del mástil siempre idéntica a la distancia del inicio (ver figura 3): «Ves, si la piedra cae ahí [al pie del mástil, *N. de la R.*], quiere decir que permanece siempre pegada al mástil cuando cae. Bueno, entonces, va a la misma velocidad que la nave. De todas formas es raro, pero debe ser eso, debe conservar su velocidad» (Sarah).

Al cabo de diez minutos de discusiones en pareja, iniciamos un nuevo debate:

> Prof: ¿Algún grupo puede proponerme una explicación que permita comprender por qué la piedra cae al pie del mástil, de la misma forma que cae al pie de la torre? ¿Sarah?
> Sarah: Nosotros pensamos que es posible si la piedra tiene siempre la misma velocidad que el barco.
> Elodie: Sí, pero cuando sale del mástil, ya no tiene velocidad.
> Sarah: Pues tiene que conservarla, si no, caería detrás.
> Prof: ¿Alguien más pensó que la piedra conservaría la velocidad de la nave? ¿Nordine?
> Nordine: Sí, nosotros pensamos que sigue a la nave a la misma velocidad que ella, porque ella le da la velocidad.

En todas las clases, algunas parejas propusieron la idea de una velocidad que se transmitiría de la nave hacia la piedra. En efecto, una experiencia del pensamiento conduce a Sarah y a su grupo a concluir que la piedra «debe conservar la velocidad» de la nave. La predicción según la cual la piedra caería al pie del mástil permite a ciertos alumnos construir una explicación cercana a la de Salviati, que no es otra cosa que una primera formulación del principio de inercia. Con el fin de reforzar a los alumnos en su enfoque, proponemos que comparen sus explicaciones con la teoría de Salviati mediante la lectura del fragmento del *Diálogo* (Anexo 1). Les recordamos que el personaje de Simplicio representa el de Aristóteles, quien inspiró el mundo académico hasta el siglo diecisiete. Después de la lectura, la mayoría de los alumnos expresan satisfacción y orgullo.

Figura 3
**Dibujo de Sarah.**

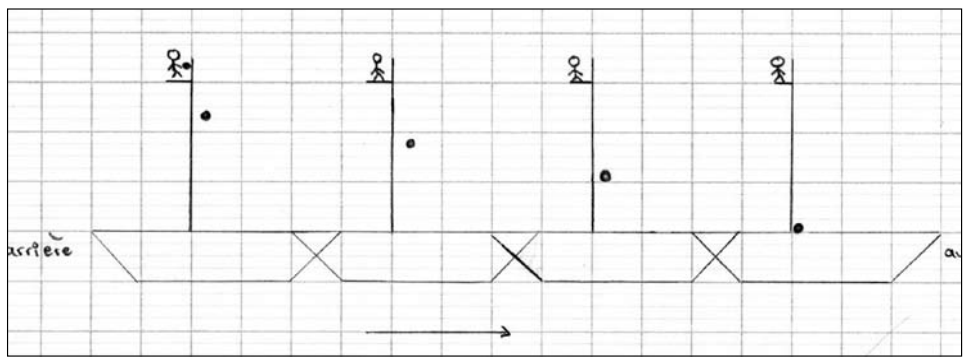

Prof: Entonces ¿cuál es la explicación propuesta por Salviati, bueno, por Galileo? ¿Linda?
Linda: Dice lo mismo que nosotros.
Sarah: Sí, somos muy buenos.
Prof: Bueno. Pero ¿qué dice?
Linda: Que el barco imprime su velocidad a la piedra, y luego es indeleblemente impresa en ella. Esto quiere decir que la velocidad no se va, aun cuando la piedra es soltada.
Prof: Sí, exactamente. Rachid ¿quieres decir algo?
Rachid: Nada. Sólo estaba pensando… ¿Esto quiere decir que somos tan inteligentes como Galileo?

Esta pregunta de Rachid y el entusiasmo de muchos alumnos muestran que la identificación con el científico es gratificante tanto para los alumnos que reconocen su razonamiento con el de Salviati, como también para aquellos que no encontraron una explicación satisfactoria. Ellos se dan cuenta de que algunos de sus compañeros son capaces de hacerlo. Algunos todavía, como Elodie, se niegan a admitir que el movimiento del barco puede ser «indeleblemente impreso en la piedra» y permanecen convencidos de que la piedra cae detrás del mástil. Las palabras de Salviati no tienen validez para estos alumnos porque «una explicación no reemplaza a una experiencia» que es la única forma de «averiguar quién tiene razón» (Elodie). Estos mismos estudiantes también lucharon para reconocer la equivalencia de las dos caídas y consideran el problema del barco como un asunto aparte. Les recordamos el experimento realizado por Gassendi en 1641 en el puerto de Marsella (Francia), pero esto sigue siendo abstracto porque los alumnos «no lo ven».

## 4. ANÁLISIS A POSTERIORI Y EVALUACIÓN

Hemos concebido nuestra trayectoria de aprendizaje basándonos en tres hipótesis: la identificación de los alumnos con el personaje de Simplicio (H1), la asociación de los dos problemas (la caída de la piedra en las dos situaciones) (H2) y la eficacia del conflicto cognitivo suscitado por el hecho de cuestionar el movimiento de la Tierra (H3).

En cuanto a la hipótesis H1, el 86% (61 sobre 71) de los alumnos implicados en nuestra investigación prevé que la piedra soltada desde la cima del mástil caerá detrás de él (ver tabla 3). Entre ellos, 52 (de 61, es decir, el 73% del total de los alumnos) toman partido de manera explícita por Simplicio, con respuestas a P1 de tipo «estoy de acuerdo con Simplicio» o «Simplicio tiene razón». Por otro lado, 8 alumnos de los 10 que respondieron que la piedra caería al pie del mástil le dieron explícitamente la razón a Salviati.

La tabla 3 indica que el 82% de los alumnos responden como lo habíamos previsto (la piedra cae detrás del mástil y al pie de la torre). Es interesante notar que el 14% de los alumnos ya formulan una respuesta que involucra un razonamiento acerca de la conservación del movimiento (los alumnos que prevén que la piedra cae al pie del mástil hablan de la «velocidad del barco» que se comunica a la piedra). Este porcentaje nos lleva a moderar la hipótesis H1 pero nos permite notar que esta situación donde los alumnos expresaron ideas opuestas favoreció un debate colectivo según las expectativas del constructivismo social.

Con respecto a la hipótesis H2, si bien los alumnos no identifican de forma espontánea la equivalencia del problema de las dos caídas, la mayoría de ellos admite que las situaciones son comparables cuando se les hace la pregunta de forma explícita. Sin embargo, nos parece importante notar que algunos alumnos siguen convencidos de que la piedra soltada desde la cima del mástil de la nave cae detrás del mismo. El cambio conceptual esperado parece, por lo menos para estos alumnos, fallar por ausencia de experimentación. También podemos cuestionar la relevancia de la identificación de las dos caídas, es decir, la semejanza entre el movimiento de rotación de la Tierra en torno a su eje y el movimiento de traslación del barco. Por ejemplo, a Jennifer le parece tan raro que la piedra caiga al pie del mástil del barco en movimiento que afirma que los problemas no son comparables ya que «la Tierra gira en círculos y la nave avanza recto». Con respecto a esto podemos suponer que hubiera sido más fácil para los alumnos admitir la equivalencia del movimiento de revolución de la Tierra alrededor del Sol con el movimiento del barco. Sin embargo, la elección que hicimos (la semejanza entre el movimiento de rotación de la Tierra en torno a su eje y el movimiento de traslación del barco) se justifica por la utilización del *Diálogo* y, a través de esto, de la estrategia de Galileo.

Tabla 3
**Repartición de las respuestas de los alumnos a las dos preguntas de las caídas libres (N = 71).**

| LA PIEDRA CAE… | N = 71 | % |
|---|---|---|
| Al pie del mástil de la nave | 10 | 14% |
| Detrás del mástil de la nave | 61 | 86% |
| Al pie de la Torre | 68 | 96% |
| Detrás de la Torre | 3 | 4% |
| *Al pie de la Torre Y al pie del mástil* | *10* | *14%* |
| *Detrás de la Torre Y detrás del mástil* | *3* | *4%* |
| *Al pie de la Torre Y detrás del mástil* | *58* | *82%* |

De hecho, nuestra intención era entregar a los alumnos un fragmento del *Diálogo* para que se enfrentaran a un episodio históricamente contextualizado. De este punto de vista, parece que la dimensión histórica de la secuencia tiene valores educativos significativos. Por ejemplo, los alumnos se ven reflejados en las respuestas de los personajes de Galileo, lo cual contribuye al debate colectivo y proporciona una cierta legitimidad a sus intervenciones. Además, después de la secuencia, algunos estudiantes expresaron su deseo de trabajar de nuevo con un método similar. Les pedimos que explicaran por qué:

> Brahim: Porque cuando uno comete un error, tenemos menos vergüenza sabiendo que los científicos cometieron errores iguales. Simplicio dijo que la piedra cae detrás del mástil. Eso también fue lo que casi todo el mundo pensaba.
> Elise: Salvo que unos de nosotros dijeron de una vez lo que dijo Galileo.
> Brahim: Sí, pero luego llegamos a cambiar de opinión, y como Galileo encontramos la explicación sin que la profesora nos ayudara. Bueno, sí, nos dijo que la piedra cae al pie del mástil. Después encontramos la explicación solos.
> Prof: Entonces, una secuencia así les gustó porque se dieron cuenta de que pensaban como un gran erudito. ¿Eso es todo? ¿Mohamed?
> Mohamed: No, también me gustó porque podemos discutir entre nosotros. A veces no estamos de acuerdo, tenemos que tratar de convencer a los demás. Como ellos [*Mohamed enseña el fragmento*] tampoco están de acuerdo.

La comparación de su razonamiento con el de Simplicio lleva al alumno a minimizar el alcance de su error y le permite darse cuenta de que equivocarse no es necesariamente malo. Además, la mayoría de los alumnos parecen conscientes de la trayectoria intelectual que les lleva a acercar la conservación del movimiento: «Llegamos a cambiar de opinión, y como Galileo encontramos la explicación» (Brahim). Esta concienciación procede de un acto metacognitivo en el que el alumno analiza las tareas realizadas, los procedimientos seguidos y los resultados obtenidos. En la frase anterior, Brahim dijo que necesitaba saber que la piedra cae al pie del mástil para seguir adelante.

Por último, con respecto a la tercera hipótesis, reconocer que el lugar donde cae la piedra debe ser idéntico en la Tierra y en la nave por su movimiento respectivo debía conducir a los alumnos a concluir que la Tierra es inmóvil. De esta incongruencia debía surgir una reevaluación de la situación de la nave (H3). Ahora bien, resulta que todos los alumnos que responden «como» Simplicio admiten que la equivalencia de las dos caídas cuestiona el movimiento de la Tierra. Algunos de ellos modifican por iniciativa propia la predicción formulada en el caso de la nave: «Tal vez nos equivocamos sobre la nave» (Nordine). Al parecer, la incongruencia puesta en escena fue productiva. Finalmente, la necesidad de que la piedra caiga al pie del mástil da pie a la construcción del principio de conservación: «Tiene que conservar [la velocidad del barco, *N. de la R.*], si no caería atrás» (Sarah). Claramente, el conflicto cognitivo subyacente a la hipótesis H3 hubiera sido posible sin las etapas 1 a 3 de la secuencia de aprendizaje, pues la mayoría de los alumnos necesitaron conocer el lugar de la caída de la piedra soltada de la cima del mástil de la nave en movimiento, lugar que les fue entregado en la etapa 4. Sin embargo, las etapas 1, 2 y 3 desarrollaron un papel por lo menos socioconstructivista, ya que les permitieron a los alumnos entrar en un proceso de debate, expresar opiniones opuestas y argumentar con el único apoyo de sus facultades racionales, ya que no tuvieron la oportunidad experimentar.

### NOTAS

1. Galileo estaba también convencido del movimiento de traslación de la Tierra. Sin embargo, parece que eligió tratar del movimiento de rotación de la Tierra en torno a su eje para no despertar las iras de la Santa Inquisición. De hecho, cuando Galileo escribió su *Diálogo*, las autoridades eclesiásticas le obligaron a expresar puntos de vista opuestos, sin tomar partido por Copérnico. Oficialmente se trata solamente de presentar los dos sistemas en el mundo con sus argumentos respectivos. En un capítulo que introduce la traducción francesa al *Diálogo*, Francisco de Gandt y René Fréreux ofrecen el siguiente análisis: «El texto es un diálogo donde diferentes puntos de vista aparecen, donde se recuerda de vez en cuando que no se trata de decidir a favor de Copérnico. De hecho, las profesiones de fe y de las protestas de neutralidad son muy formales y superficiales» (Gandt y Fréreux, 1992, 3).
2. Las razones que parecen haber conducido a Galileo a tomar esta decisión se explican en la nota 1.

**REFERENCIAS BIBLIOGRÁFICAS**

ARTIGUE, M. (1998). Ingeniería didáctica, en Artigue, M., Douady, R., Moreno, L., Gómez, P. (eds.). *Ingeniería didáctica en educación matemática*. Colombia: Una empresa docente.

AUSUBEL, D.P., NOVAK, J. y HANESIAN, H. (2000). *Psicología educativa: un punto de vista cognoscitivo*, 2.ª ed., Trillas: México.

CHEVALLARD, Y. (1991). *La transposición didáctica: del saber sabio al saber enseñado*, Argentina: Aique.

CLEMENT, J. (1982). Students' Preconceptions in Introductory Mechanics, *American Journal of Physics*, 50, pp. 66-71.

DOISE, W. (1991). La estructuración cognitiva de las decisiones individuales y colectivas de adultos y niños. *Anthropos: Boletín de información y documentación*, 27, pp. 6-12.

DE HOSSON C. y KAMINSKI W. (2007). Historical Controversy as an Educational Tool. Evaluating Elements of a Teaching-learning Sequence Conducted with the «Dialogue on the Ways that Vision Operates». *International Journal of Science Education*, 29(5), pp. 617-642.

GALILEI, G. (1995). *Diálogo sobre los dos máximos sistemas del mundo*, Alianza.

GALILI, I. y HAZAN, A. (2001). The Effect of a History-Based Course in Optics on Students' Views about Science', *Science & Education*, 10(1-2), pp. 7-32.

GARRISON J.W. y LAWWILL K.S. (1993). Democratic Science Teaching : a Role for History of Science, *Interchange*, 24(1-2), pp. 29-39.

GIL-PÉREZ, D. y CARRASCOSA, J. (1990). What to do about science misconceptions? *Science Education*, 74(4).

DE GANDT, F. y FRÉREUX, R. (1992). Introduction au *Dialogue sur les deux grands systèmes du Monde* de Galilée, Seuil: París.

GUEDJ, M. (2005). Utiliser des textes historiques dans l'enseignement des sciences physiques en classe de seconde des lycées français: compte-rendu d'innovation. *Didaskalia*, 26, pp. 75-95.

HALLOUN, I.A. y HESTENES, D. (1985). Common Sense Concepts about Motion, *American Journal of Physics*, 53, pp. 465-467.

HARRES, J.B.S. (2005). La física de la fuerza impresa como referente para la evolución de las ideas de los alumnos, *Enseñanza de las Ciencias*, Número Extra, VII Congreso, pp. 1-5.

JOSHUA, S. y DUPIN, J.M. (2005). *Introduccion a la didáctica de las matemáticas y de las ciencias*.

KAPLAN, A. (1992). Galileo: An Experiment in Interdisciplinarity Education, *Curriculum Inquiry*, Blackwell publishers: Toronto, pp. 255-288.

LIMON, R.R. (2004). *Historia de la psicología y sus aplicaciones*, México.

MORA, C. (2009). Una revisión sobre ideas previas del concepto de fuerza, *Latin American Journal of Physics Education*, 3(1), pp. 72-86.

MATHY, P. (1997). *Donner du sens aux cours de sciences. Des outils pour la formation éthique et épistémologique des enseignants*. Bruxelles: De Boeck.

MATTHEWS, M. (2000). *Time for Science Education, how Teaching the History and Philosophy of Pendulum Motion can Contribute to Science Literacy,* Kluwer Academic/Plenum Publishers: Nueva York.

MCDERMOTT, L.C. (1984). Research on Conceptual Understanding in Mechanics, *Physics Today*, pp. 24-32.

MONK, M. y OSBORNE, J. (1997). Placing the History and Philosophy of Science on the Curriculum: A Model for the Development of Pedagogy. *Science Education*, 81(4), pp. 405-424.

PICQUART, M. (2008). ¿Qué podemos hacer para lograr un aprendizaje significativo de la física? *Latin American Journal of Physics Education*, 2(1), pp. 29-35.

POSNER, G.J., STRIKE, K.A., HEWSON, P.W. y GERTZOG, W.A. (1982). Accommodation of a scientific conception: Towards a theory of conceptual change. *Science Education*, 66(2), pp. 211-227.

SALTIEL, E. y MALGRANGE, J.L. (1980). Spontaneous Ways of Reasoning in Elementary Kinematics, *European Journal of Physics,* 1(73),

SALTIEL, E. y VIENNOT, L. (1985). ¿Qué aprendemos de las semejanzas entre las ideas históricas y el razonamiento espontáneo de los estudiantes?, *Enseñanza de las Ciencias*, 3(2), pp. 137-144.

SANMARTÍ, N. y CASADELLA, J. (1987). Semejanzas y diferencias entre las concepciones infantiles y la evolución histórica de las ciencias: el ejemplo del concepto de fuerza y especialmente del de fuerza de gravedad. *Enseñanza de las ciencias*, 5(1), pp. 53-58.

SEBASTIA, J.M. (1984). Fuerza y movimiento: la interpretación de los estudiantes, *Enseñanza de las Ciencias*, 2, pp. 161-169.

SLISKO, J. (2008). La historia de la física en la enseñanza. Desde los objetivos curriculares hasta la práctica docente. *El Cronopio*, pp. 16-21.

VIENNOT, L. (1979). *Le raisonnement spontané en dynamique élémentaire*, París: Hermann.

VIENNOT, L. (2002). *Razonar en física. La contribución del sentido común*. Madrid: Ant. Machado libros (Visor Distribuciones).

**Anexo 1**

**Fragmento del *Diálogo sobre los dos máximos sistemas del mundo: ptolemaico y copernicano* de Galileo Galilei (1632)**

**Salviati:** Más bien deseo que continuéis en él y que tengáis por seguro que el efecto de la Tierra deba responder al de la nave, a condición de que si eso se mostrase perjudicial para vuestras necesidades, no se os ocurra cambiar de idea. Vos decís: puesto que si la nave está quieta, la piedra cae al pie del mástil, y si está en movimiento cae lejos del pie, por la inversa, del hecho de que la piedra caiga al pie se infiere que la nave está quieta, y del hecho de que caiga lejos se deduce que la nave se mueve. Y puesto que lo que ocurre en el caso de la nave debe igualmente suceder en el caso de la Tierra, del hecho de que la piedra caiga al pie de la torre se infiere necesariamente la inmovilidad del globo terrestre. ¿No es éste vuestro razonamiento?
**Simplicio:** Es exactamente así, resumido de un modo que lo hace facilísimo de entender.
(...)
**Salviati:** Muy bien. ¿Habéis hecho alguna vez la experiencia de la nave?
**Simplicio:** No la he hecho. Pero creo que los autores que la aducen, la han observado diligentemente. Por lo demás, la causa de la diferencia se conoce tan claramente que no deja lugar a dudas.
**Salviati:** Vos mismo sois un buen testimonio de que es posible que los autores la aduzcan sin haberla hecho, puesto que sin haberla hecho la tenéis por segura y os remitís a la buena fe de su afirmación. Así también no sólo es posible, sino necesario, que también ellos hayan hecho lo mismo, quiero decir remitirse a sus antecesores, sin llegar nunca a alguien que la haya hecho. Porque cualquiera que la haga hallará que la experiencia muestra todo lo contrario de lo que se ha escrito. Esto es, mostrará que la piedra cae siempre en el mismo lugar de la nave, tanto si está quieta como si se mueve con cualquier velocidad. Por lo que, por ser la misma la argumentación referente a la Tierra que a la nave, del hecho de que la piedra caiga siempre perpendicularmente al pie de la torre no se puede inferir nada sobre movimiento o reposo de la Tierra.
(...)
**Simplicio:** Vos queréis decir como última conclusión que, al moverse la piedra con un movimiento indeleblemente impreso en ella, no dejará la nave, al contrario la seguirá y finalmente caerá en el mismo lugar en que cae cuando la nave está quieta. Y así digo yo también que sucedería si no hubiera impedimentos externos que estorbasen el movimiento de la piedra después de haber sido dejada en libertad. Dichos impedimentos son dos. Uno consiste en que el móvil es impotente para romper el aire sólo con su ímpetu, faltándole el de la fuerza de los remos, del cual era partícipe, como parte de la nave mientras estaba sobre el mástil. El otro es el movimiento nuevo de caer hacia abajo, que necesariamente constituye un impedimento para el movimiento hacia adelante

**Anexo 2**

**Cuestionario distribuido a los alumnos en la etapa 1**

Imagine que alguien deja caer una piedra desde el último piso de una torre. ¿Dónde caerá la piedra?
  – ¿Al pie de la torre?
  – ¿Detrás de la torre?
  – ¿Delante de la torre?
Justifique su respuesta.

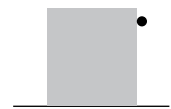

Imagine ahora que alguien deja caer esa misma piedra desde la punta del mástil de una nave en movimiento uniforme sobre un mar perfectamente quieto. ¿Dónde caerá la piedra?
  – ¿Al pie del barco?
  – ¿Detrás del barco?
  – ¿Delante del barco?
Lea el siguiente texto y escriba lo que piensa al respecto.

**Simplicio**: tenemos la experiencia muy adecuada de la piedra dejada caer desde la cima del mástil de la nave que, si la nave está quieta, cae al pie del mástil, pero si la nave avanza, cae tan lejos del mismo punto cuanto la nave, en el tiempo de la caída de la piedra, se ha desplazado hacia adelante. Y no son pocas brazas cuando el avance de la nave es veloz.
**Salviati**: Yo sin experiencia estoy seguro de que la piedra caerá al pie del mástil.

NB: aclaremos que la piedra es lo suficientemente pesada para que el efecto del aire sea omitible. Recordemos además que la Tierra gira sobre su propio eje en razón de una vuelta cada 24 horas, lo cual equivale, para un punto situado sobre la superficie de la Tierra (en París, por ejemplo), aproximadamente a 260 metros por segundo.

# Using a historical controversy in a learning context: the case of a «didactic engineering» elaborated from Galileo's «*Dialogue Concerning The Two Chief World Systems*»

DE HOSSON, CÉCILE
Laboratoire de Didactique André Revuz, Université Paris Diderot
cecile.dehosson@univ-paris-diderot.fr

## Summary

It's commonly admitted in the Science Education Research community that the use of history of science, can provide students with a more comprehensive view of science and in some cases helps them to get a better appropriation of scientific concepts and laws. One the stake of such way of teaching consists in avoiding the reductive and anecdotic form that can be found in most of textbooks. In that perspective, the strategy consisting in the transposition of a historical problem inside the science classroom opens a fecund way for an educational use of the history of science. In this paper we present a classical Mechanics teaching sequence elaborated from Galileo's *Dialogue Concerning The Two Chief World Systems* (1632). This sequence aims to learn the principle of the conservation of the movement. The elaboration and the analysis of the sequence take place in the 'Didactic Engineering' framework that allows the extent of the impact of an educational action. It consists in confronting an *a priori* analysis where local hypothesis underpinning the researcher's actions are involved to an *a posteriori* analysis leaning on the data coming from the effective performed sequence.

Students' ideas in Newtonian mechanics have been the target of a wide current of researches all over the world. In has been shown that most of the students think that the movement of a body implies a cause that is the action of a motor. This type of reasoning leads the students to predict that a stone released by somebody at rest in a moving support will fall down behind this person. In other words, in the absence of a physical link between a released object and its previous support, the persistence of the motion of the support in the object is denied. This echoes the historical ideas as implemented by Galileo in his *Dialogue*. Thus, we benefit this proximity to elaborate our sequence on the basis of some pieces of the *Dialogue*. Moreover, we favor a socio-constructivist approach where the confrontation of various points of views takes part of the learning process by implementing a controversial question chosen to be the starting point of a collective debate in the classroom. The organization of the sequence is as follows: in the first step of the sequence, the students are invited to predict where a stone released from the top a tower on the one hand should fall, and on the other hand, from the top of a mast of a boat keeping a constant velocity. They are also invited to read a small piece of Galileo's *Dialogue* where Simplicio and Salviati discuss the place where a stone released from the top of the mast of a ship should fall. We expect the student advocate for one of the answers staged in the small piece. The second step aims at making the students admit the equivalence of the two falls (here the effects of the Coriolis force and that of entertainment are neglected). Our hope is to lead them to a weird conclusion underpinned by the following reasoning: if the two falls are equivalent, then the stone released from the tower should fall behind it; if it is not the case, that means that the Earth is quiet. Facing this conflictive situation, the students should promote a new conclusion for the falling place of the stone released from the mast of the boat (step 3). This will happen thanks to a reasoning close to the expected one: the stone keep the horizontal velocity of the boat even when the physical link between it and the mast does not exist anymore (step 4). Another piece of *Dialogue* is then provided to the students where Salviati explains what could be considered as the first form of the inertial principle.

The teaching sequence was performed with three groups of more or less 25 fifteen-year-old students each one lasted two hours. The whole sequences have been videotaped and transcribed. Our analysis consists in the proving of the following hypothesis (Hn) that raised from the *a priori* analysis: 1) The identification of the students with the character of Simplicio (H1), 2) The acceptance by the students of the equivalence of the two falls (H2), 3) The efficiency, in term of conceptual change, of the cognitive conflict provoked by the incongruity concerning the immobility of the Earth (H3). Many students did not so easily accept the equivalence of the two falls. Nevertheless, the students who accepted this equivalence reconsidered their prediction concerning the place of the fall of the released stone in the case of the moving boat. This change in the students reasoning was motivated by the fact that there was no doubt for them that the stone released from the top of a tower falls down at the bottom of this tower, whereas the Earth rotates around its axis. This led these students to elaborate a new operative reasoning in order to explain the falling place of the stone in both cases. In this new operative reasoning, the velocity of the moving body (the tower or the mast) is transmitted into the released stone so that it can «follow» the moving body (the tower o the mast).